\documentclass[conference]{IEEEtran}
\ifCLASSINFOpdf
   \usepackage[pdftex]{graphicx}
\else
\fi
\usepackage{multirow}

\usepackage{url}

\begin{document}
\title{UTPO: User's Trust Profile Ontology - Modeling trust towards Online Health Information Sources}

\author{\IEEEauthorblockN{Prakruthi Karuna, Hemant Purohit, Vivian Motti}
\IEEEauthorblockA{Volgenau School of Engineering\\
George Mason University\\
Fairfax, Virginia 22030\\
Email: \{pkaruna, hpurohit, vmotti\}@gmu.edu}}


%Telephone: (800) 555--1212, Fax: (888) 555--1212}


\maketitle

\begin{abstract}
Despite the overwhelming quantity of health information that is available online today, finding reliable and trustworthy information is still challenging, even when advanced recommender systems are used. To tackle this challenge and improve our recommended sources, we need to first understand and capture the user behavior of what is considered to be trustworthy. This paper presents a taxonomy of relevant factors that influence user' trust towards online health information sources. We design a survey experiment to validate the taxonomical factors, and propose an ontology using the taxonomy of factors, such that each user's trust could be modeled as an instance of our ontology and this could later be used programmatically to recommend trustworthy information sources. Our work will inform the design of  personalized recommendation systems and websites to improve online delivery of health information.

\end{abstract}


% \fi
\IEEEpeerreviewmaketitle

\section{Introduction}

Seeking health information and advice from medical providers is expensive and time consuming and sometimes, the medical problem and associated symptoms may not be severe as per a user perception. As a result, an increasing number of patients with varying intents seek health information on the Internet [3,4]. Despite being extensive, the health information available online varies in quality, and sifting the enormous web for good quality information can be overwhelming for users. If a set of trustworthy information sources were available, this overwhelming task would be simpler. As per dictionary definition, trust is a \emph{'belief that someone or something is reliable, good, honest, effective, etc.' }Therefore, it is critical to understand what fundamental factors contributes in building such a belief in the online world. \\

A user trusts a website when his/her internal attitude (or implicit requirements) towards the website matches with the website [1]. Internal attitudes and rationale for justifying trust towards a website vary individually per user and his/her search intentions [3]. For example, while a user trusts a mechanic to repair his/her car, he/she will not trust him/her for medical advice. In the health domain, intentions include: acquiring knowledge, evaluating a provider, or even sharing personal experiences in exchange for empathy; for each of these intentions, the underlying rationale to trust can vary. For a website to be considered trustworthy, it has to be in accordance with the user's internal attitude at the time of a query. Trust is also formed based on the user experience after several interactions with a website depending on the situational cues received. These cues or factors later determine the value of trust for a given situation or context.\\

Trust in a health information source is highly subjective to the user, his/her health status and intent of information seeking. Unlike other domains, users seeking health information are not neutral processors but they tend to have strong initial expectations and prefer certain information types [2]. Therefore, our key research questions are the following:\\
\begin{enumerate}
\item What are the factors for a user to trust a health information source on the Internet?
\item Can we design a comprehensive meta-view of trust factors for web information sources in the healthcare domain?
\end{enumerate}

Prior research has investigated different trust factors for users of online sources, however, a meta-view of all such factors is lacking. To the best of our knowledge, no experimental studies have been conducted so far to analyse trustworthiness from a user-behavior perspective in the context of health information seeking in online sources, especially envisaging the design of recommender systems. Furthermore, within the context of health information seeking, there is still a need to validate relevant trust factors priorly studied in the general search context. In this paper, we propose a taxonomy of trust factors that are relevant to identify trustworthy and reliable health based information sources, using this taxonomy, we built a consolidated knowledge source (Ontology). This ontology can be adapted to an individual level to define and model trust requirements for a given user. Our trust model also associates the user's trust to his/her intent; thus allowing a single user to be associated with multiple types of trust (i.e. one type of trust per intention).\\

Our approach consists in first reviewing related work on trust factors in online medium to define a taxonomy of trust factors, as Section 2 and 3 present. We then design an online survey to understand users' perspectives on these factors and to validate our meta-view model of trust factors, as Section 4 describes. Finally, we present the survey results and our analysis in Section 5. Based on our acquired knowledge from related work and the survey, in section 6, we present the ontology of user's trust profile (named UTPO).

\section{Related Work}
We discuss prior literature in three areas that are relevant to formulate the foundation for UTPO.

\emph{Health Information Sources} - Search engines like Google are often used to search for information on the web, however those searches are often limited to keyword queries, and it is challenging to identify user's intent and therefore, to provide reliable information sources as results [3]. On the other hand, a new form of medium is the social media based search. Jadhav et. al [18] developed a system that searches through near real-time Twitter data to extract relevant and reliable health information using semantic techniques that enhance the search query. This system is also limited though, as it focuses only on Twitter data.
\emph{Human Factors in Website Interactions} -Identifying factors and modeling trust is an active research area. Existing models define trust based on different subsets of factors [2,4,7,8]. It is challenging to know which subset of factors is the most suitable for a user, especially in particular contexts of use.\\ 
The main shortcoming of studies in this area is that, they have studied the factors on inconsistent information sources belonging to different domains, making it unclear whether these factors are all applicable to health domain as well. Also, if there is a need to investigate if such factors have a correlative effect on the understanding of user's perception of a website.  
\emph{Trust Ontology} - Ontologies based on trust models for business systems [5], semantic services-oriented environment using general trust concepts [6] have been developed. However, these do not apply specifically to health information.\\
For the ultimate motivation of this research for supporting the design of a recommendation system, we note that a true trustworthy recommendation is only possible when the recommendations can be adapted to both implicit  (user's internal attitude) and explicit  (keywords for search) requirements of the user. All the aforementioned studies of human factors have pruned out a major chunk of factors to leave behind only a few of the common factors. For a recommender system to adapt itself to provide trustworthy recommendations for non-neutral information seeking users at an individual level, it should have the knowledge base of various trust factors as well as, how they impact the user's trust and their mutual influences on user's trust, and any relationships to other factors.

\section{Taxonomy of Factors to Trust and Ontology Class Description}

Based on our review of prior research work, we formally define a taxonomy of trust factors for health information and then validate it via a user survey (discussed in section 4). Our proposed taxonomy (Figure 1) includes five categories: Social, Visual, Security and Privacy, Information Quality, and User's Personality.

\begin{figure*}[ht!]
\centering
\includegraphics[height=50mm, width=140mm]{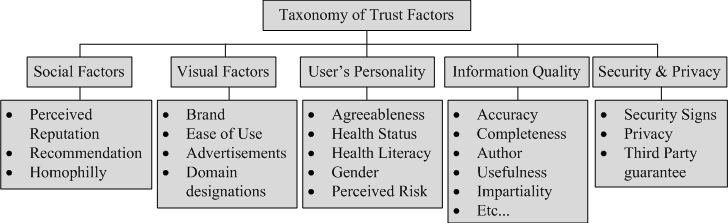}
\caption{Taxonomy of sixteen Trust Factors for Health Information in Online Sources \label{overflow}}
\end{figure*}

\subsection{Social factors}
\subsubsection{Perceived Reputation}

a positive reputation (e.g., when the online source provides useful information for the user) leads to higher chances of trust [5,6]. A positive reputation for a website can be realized in different ways, including the analysis of the users' comments (when those are available). The "Perceived\_Reputation" class holds the minimum value of positive reputation a user expects from a website using its slot "threshold" ranging {0-1}. 

\subsubsection{Recommendation by family, friends or other trusted users}
A positive reference to a website increases the possibility of trust. Conversely, a negative reference reduces the trust [3]. A "Recommendation" class defines two boolean faceted slots "requirePositiveReco" and "ignoreNegativeReco" to model the user's opinion concerning the recommendations.

\subsubsection{Homophily}
drives credibility for health information presented in a discussion group. Participants who perceive higher levels of homophily give higher evaluations [9], leading to greater likelihood to act on the advice, and thus implying higher trust levels [8]. The "Homophily" class defines a "threshold" property ranging from {0-1}, it refers to the minimum value of commonality between user and author.

\subsection{Visual factors}

\subsubsection{Brand}
Presence of familiar or trusted brand logos can increase the trust on a website [7]. "Brand" class defines a Boolean faceted slot "isBranded" that holds a value '1' if the user is interested in websites built by known brands and a value '0' otherwise.

\subsubsection{Ease of Use: Professional layout and writing, responsiveness, and availability}
A website viewed as easily operable, available and responsive is likely to be adopted and trusted by users [7,10,11]. Within a thirty-second window, participants are able to efficiently sift information ? recognizing and rejecting general portals and sales sites [8]. "Ease\_of\_use" can be measured by the number of clicks and scrolls required to get the information of interest on a given website. "isEasy" predicate holds the maximum  (zero or more) number of navigations a user is willing to perform to find the information. 

\subsubsection{Advertisements and domain designations}
Presence of ads can reduce trustworthiness of websites; for example, when a website is sponsored by a drug company and presents extremely good reviews about its products, it could raise questions about the trustworthiness of those reviews and thus the website. Ads have positive effects on websites with .com, and .edu domain designations and deleterious effect on websites with .org; and websites on .gov are considered the most trustworthy [12]. "allowAds" is a boolean variable in "Advertisements" class that holds "1" if the user is interested in websites sponsored by ads and "0" otherwise. The "Domain\_Designation" class defines a variable "preferedDomains" that holds a list of strings, that the users deem to be trustworthy.

\subsubsection{Prior experience and Personalization}
can both indicate and influence trust. Good personalization requires information about the user and providing personal information can suggest that the user trusts the website. Similarly, a returning user could indicate an inherent trust in the website [7,8,10]. "preferVisitedSites" is a boolean variable in "Prior Experience" class that holds "1" if the user is interested to re-visit a website and "0" otherwise.
\subsection{Security \& Privacy}

\subsubsection{Third Party guarantee}
If a trusted third party guarantees the health information, it could be trustworthy. The Health on the Net (HON) Foundation Code of Conduct [13] and the Truste [14] are some of the more commonly occurring health related seal programs. If a user understands and trusts the programs, s/he is likely to trust the web sites bearing the seals [10].
"requireSeal" is an attribute of "Third\_Party\_Guarantee" class that holds a list of guarantees the user would like to see on a website.

\subsubsection{Visible security sign}
The presence of security signs (e.g., lock symbol) enhances its trustworthiness [7]. "encryption" is a boolean variable in "Visible\_Security\_Sign" class with value "1" if user wishes to see only websites that have encryption enabled and "0" otherwise.

\subsubsection{Privacy}
Users of online health information are generally skeptical about sharing their health information online, as such information could get misused. To put users at ease, websites posts their privacy policy. However, the effectiveness of these policies in mitigating user's perceived risk depends on the user's knowledge and trust in them [16].
"displayPolicy" is a boolean variable in "Privacy" class with value "1" if user wishes to see only websites that lists their privacy policy and "0" otherwise.

\subsection{Information Quality (IQ)}
IQ is the most important  trust factor, being directly proportional to trustworthiness of web page. As several researchers identified, IQ is a complex concept, involving various factors and quality criteria; Jayawardene et al [15] provides a consolidated list of dimensions for IQ, below are a few:

\subsubsection{Reliable content} is the belief that the other party will keep his or her word, fulfill promises, and be sincere. For websites, this means that there are no false statements and the information on the site is correct.[22]

\subsubsection{Competence}
means that the website has the resources (whether information, technical, financial, or human) and capabilities needed for the successful completion of the action and the continuance of the relationship [23]

\subsubsection{Usefulness} The extent to which the user is informed by and can make use of the information [22].

\subsubsection{Proven expertise \& authority} or a knowledgeable source of information is identified as one of the top trust markers. [22]

\subsubsection{Open and available cases or conversations} showing success stories for the user could improve users willingness to provide more personal information and increase trust.
The selection of quality criteria is modeled as attributes for the "Information\_Quality" class and depends on the context of the application that is using the UTPO ontology.

\subsection{User's Personality}
Trust on a website also heavily depends on the user's personality, and therefore, following factors can be related to the user's trust. In our UTPO ontology, we model user as a class of six attributes. These attributes (defined next) are helpful to compare different users' profiles.

\subsubsection{Agreeableness}
Agreeableness is a trait related to the likeability and social conformity of individuals, a person with an agreeable nature tends to trust websites more often [16]. "Agreeableness" attribute holds a value in the range of {0-1}.

\subsubsection{Health Status}
Users who believe to be in "good" health condition tend to trust health websites, while those with poor health status rely more on information given by their healthcare provider [9], [8]. "hStatus" holds a string value from the set {Good, Average, Poor}.

\subsubsection{Health literacy}
Online information is more accredited and trusted if the user already has knowledge about the health topic [16]. "hLiteracy" holds a string value from the set {Good, Average, Poor}.
\subsubsection{Gender and Age}
on average, female users placed greater trust in health websites [9]. Younger consumers trust  more on health websites provided by personal doctors, medical universities, federal governments and local communities [15]. Although a strong correlation between age and trust towards a website seems to exist, the usage of specific ages or age groups as a trust predictor is unclear.  An optional "gender" and "age" attributes could be used based on the context.

\subsubsection{Perceived Risk}
refers to the extent to which a user views the consequences of acting on the health information provided in a website to be uncertain [11]. The higher the uncertainty, the lower the trust. "perceivedRiskLevel" attribute of "Perceived Risk" class tracks the risk level using a value from the set {high, medium. Low}

\section{Experimental Design}
To understand how users access, use and trust health information websites, we designed and conducted a web-based survey using Google forms to understand how users access, use and trust health information websites. Our survey design consists of three parts. \\
In Part 1, to understand their intent, we asked our recruits questions about their search patterns, for example: "Have you ever searched for health information online?" and "Why have you looked for health information online?" Based on their responses, we learned that they had varied intents, including to: "Evaluate a provider", "Acquire knowledge", "Self-diagnose", "Find home remedies" and "Get a quick answer".\\
In Part 2, we provided our recruits with a list of websites and asked them related questions to understand how they would evaluate these websites. Table 1 lists the websites selected, the rationale for selecting these websites, as well as the corresponding questions. We also asked our recruits to rank these five websites according to their trustworthiness.\\
Seeking to understand the effects of factors on user's justification to trust, in Part 3, we provided our recruits with a set of 12 statements (Table 2) and asked their opinion. A 5-point Likert scale including 'Strongly Agree', 'Agree', 'Undecided', 'Disagree', and 'Strongly Disagree', was used to that end. \\
Unlike other research studies, our survey focuses on: (a) a comprehensive set of factors, (b) the correlation between those factors, and (c) the specificities of the health information domain. The depth and breadth of our survey guided our recruits to think beyond the theoretical definitions of trust factors, by employing an empirical decision-making approach, the results give us a distinct and valuable data set.

\begin{table*}[ht]
\caption{Websites chosen for Part 2}
\centering
\begin{tabular}{p{2cm}p{4cm}p{6cm}}
\hline
Websites & Rationale &  Questions\\ [0.5ex] % inserts table %heading
\hline
WebMD&Recommended in CAPHIS trustworthy websites&Information provided was relevant and useful for me\\
&&Website is easy to use (loads quickly, easy to find information)\\
&&This website has a good reputation\\
&&Tips provided to cure headache where risky to implement\\
Medline Plus&Government sponsored website&Information provided was relevant and useful for me\\
&&Website is easy to use (loads quickly, easy to find information)\\
&&I trust this website because it is owned by a government\\
&&Tips provided were risky to implement\\
Everyday health & Shows a video of a provider addressing the website's users.&Information provided was relevant and useful for me \\
&&Website is easy to use (loads quickly, easy to find information)\\
&&I trust this website because it has a Dr. addressing it\\
Cancer Survivors Network   &Health-based forum were users could share their health concerns.&Information provided was relevant and useful for me\\
&&I'm concerned about my privacy while using this website\\
&&I know this website has a good reputation\\
&&I trust this website because it is written by people who have experienced health conditions\\
Breathe in and out&Website marked as Advertisement by Google. &Website was easy to use, understood what they had to offer.\\
&&I did not download their product as I believed it could be malicious, a third party guarantee could help.\\
&&I don't trust this website because it is advertising a product\\
&&Tips provided are risky to implement\\[1ex]
\hline
\end{tabular}
\label{table:nonlin}
\end{table*}

\begin{table*}[ht]
\caption{List of statements used in Part 3}
\centering
\begin{tabular}{l l}
\hline\hline
No. & List of Statements \\ [0.5ex] % inserts table %heading
\hline
1&Quality of information affects website's trustworthiness  \\
2&I would trust a website if it is Recommended to me by my family or friends \\
3&I would trust a website if it has a good reputation \\
4&If a website is sponsored by a brand, I would trust it\\
5&If the author has experienced similar health conditions as me, then I would trust it\\
6&Website needs to load quickly and be easy to use.\\
7&Websites that contain advertisements are not trustworthy\\
8&I would trust a website if it prescribes a lower risk cure. \\
9&I would trust a website if I can personalize it.\\
10&If I am re-visiting a website, then I would trust it. \\
11&A third party guarantee enhances website's trustworthiness \\
12&Visible security sign, clear privacy policy increases website's trustworthiness\\
[1ex]
\hline
\end{tabular}
\label{table:nonlin}
\end{table*}

\section{Results and Analysis}
We created our survey using Google Forms, then to collect the responses, we disseminated it to fifteen users by email. Nine responses were received in total, from a gender-balanced participants' sample with intermediate to advanced computer skills. 

As our recruits were asked to order the websites in reducing trustworthiness, we picked the top three websites that each of our recruits selected, symbolizing the ability of a user to visit fewer websites than the recommended. Figure 2 shows the frequency of websites chosen.

The results obtained for Part 3 are shown in Figure 3. From the results we noted that no two recruits had the same opinion on factors, either they selected different sets of factors or prioritized the factors differently. This informs about the varying user intent for the information seeking as well as source trustworthiness. 

\begin{figure}[!t]
\centering
\includegraphics[width=3.5in]{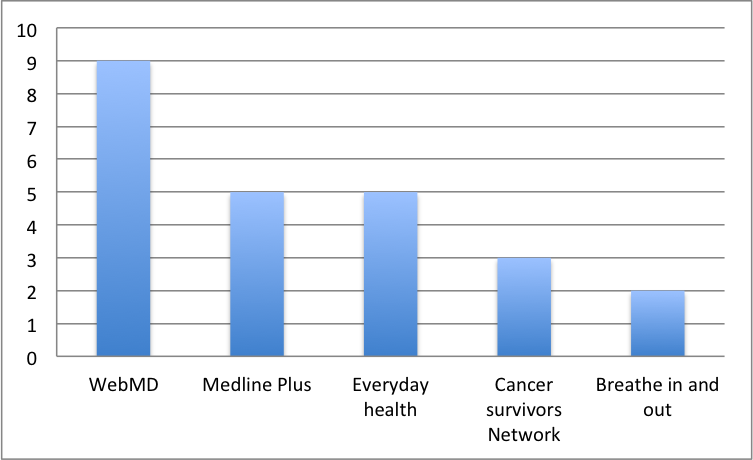}
% where an .eps filename suffix will be assumed under latex, 
% and a .pdf suffix will be assumed for pdflatex; or what has been declared
% via \DeclareGraphicsExtensions.
\caption{Frequency of websites selected based on trustworthiness}
\label{fig_sim}
\end{figure}

\begin{figure}[!t]
\centering
\includegraphics[width=3.5in]{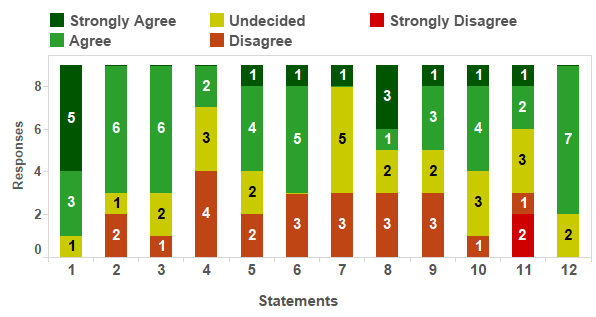}
% where an .eps filename suffix will be assumed under latex, 
% and a .pdf suffix will be assumed for pdflatex; or what has been declared
% via \DeclareGraphicsExtensions.
\caption{Experiment Results from nine users from part 3}
\label{fig_sim}
\end{figure}

%\begin{figure}[ht!]
%\centering
%\includegraphics[height=2in, width=3in]{t1.png}
%\caption{Experiment Results} \label{overflow}
%\end{figure}

\section{Ontology Design}
As a result of the factors identified in the related works and our survey results, we modeled an ontology named "User Trust Profile Ontology (UTPO)" [19]. UTPO is defined in OWL (Web Ontology Language), based on the guidelines provided by Stanford [20]. UTPO defines trust factors as classes and their relation to the user, facilitating interoperability between users and information sources.
The structure of UTPO is shown in Figure 2. Each of the classes representing factors is a subclass of "Influencing factors" class. This class defines an attribute called "weight" with value in the range {0-1} that defines the importance of each factor when computing trustworthiness of a website. A user class could also have associations of type "friend", or "family" with other user classes.

\begin{figure*}[!t]
\centering
\includegraphics[width=4.5in]{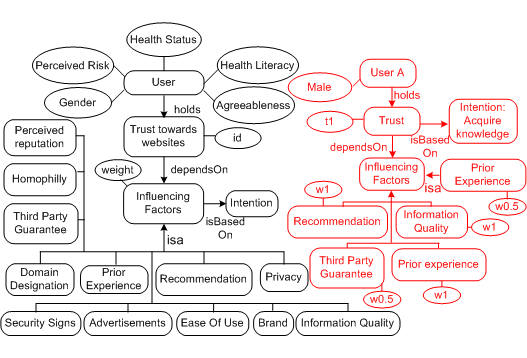}
% where an .eps filename suffix will be assumed under latex, 
% and a .pdf suffix will be assumed for pdflatex; or what has been declared
% via \DeclareGraphicsExtensions.
\caption{User Trust Ontology and an instance (user?s trust profile) }
\label{fig_sim}
\end{figure*}

%\begin{figure}[ht!]
%\centering
%\includegraphics[height=40mm, width=90mm]{UTPO.png}
%\caption{User Trust Ontology and an instance (user?s trust profile) \label{overflow}}
%\end{figure}

As stated above, every individual's justification to trust is different from one another, indicating that the trust model should be adaptable at individual levels. This could be achieved simply by instantiating the UTPO for each user. A separate instance of the trust class should be created for each of the user's intents thereby linking a single user class with multiple trust classes.
As an example, an instance of UTPO is created for user "A" from our experiment in Fig 2.  We model user's trust by considering factors that the user agrees (Strongly Agree and Agree) and would like to see in a website. We assigned a weight of "0.5" for "Agree" and "1" for "Strongly Agree".  The UTPO ontology is available with url https://goo.gl/dWtMR9.

\section{Conclusions and Future Work}
We presented a taxonomical organization of sixteen trust factors for users seeking health information online. Some factors had already been remarked in prior studies, however, those studies focused on different domains, and were not specific to online health information. To validate our taxonomy and to understand the users' perspectives on the trust factors  identified, as well as  its applicability to health information, we also conducted a user experiment. 
Using the factors and an understanding of human trust, we provide a multidimensional framework to dynamically model trust for health information websites. Our model captures the user' expectations to form justified trust. It can also adapt to changing intentions of a user. Our model can assist in: designing recommending or personalizing websites, comparing individuals, and improving collaborative filtering algorithms.
By factoring latent user attributes, such as personality and social and visual perception implications, our ontology contributes to the design and implementation of next-generation recommender systems, supporting users in finding trustworthy health information from online sources.

\end{document}